\documentclass[]{ws-rv9x6} 
\usepackage{ws-rv-har}    
\usepackage{ws-rv-thm}    
\usepackage{subfigure}    
\usepackage{enumitem}
\usepackage[percent]{overpic}
\usepackage{xcolor}
\makeindex

\begin{document}		
\newcommand{\ltsima}{$\; \buildrel < \over \sim \;$}
\newcommand{\lsim}{\lower.5ex\hbox{\ltsima}}
\newcommand{\gtsima}{$\; \buildrel > \over \sim \;$}
\newcommand{\gsim}{\lower.5ex\hbox{\gtsima}}
\newcommand{\bra}{\langle}
\newcommand{\ket}{\rangle}
\newcommand{\lprime}{\ell^\prime}
\newcommand{\lpp}{\ell^{\prime\prime}}
\newcommand{\mprime}{m^\prime}
\newcommand{\mpp}{m^{\prime\prime}}
\newcommand{\ci}{\mathrm{i}}
\newcommand{\dd}{\mathrm{d}}
\newcommand{\veck}{\mathbf{k}}
\newcommand{\vecx}{\mathbf{x}}
\newcommand{\vecr}{\mathbf{r}}
\newcommand{\vecv}{\mathbf{\upsilon}}
\newcommand{\vecw}{\mathbf{\omega}}
\newcommand{\vecj}{\mathbf{j}}
\newcommand{\vecq}{\mathbf{q}}
\newcommand{\vecl}{\mathbf{l}}
\newcommand{\vecn}{\mathbf{n}}
\newcommand{\lm}{\ell m}
\newcommand{\that}{\hat{\theta}}
\newcommand{\thatp}{\that^\prime}
\newcommand{\chip}{\chi^\prime}
\newcommand{\hs}{\hspace{1mm}}
\newcommand{\nar}{New Astronomy Reviews}
\def\gsim{~\rlap{$>$}{\lower 1.0ex\hbox{$\sim$}}}
\def\lsim{~\rlap{$<$}{\lower 1.0ex\hbox{$\sim$}}}
\def\Msun {\,\mathrm{M}_\odot}
\def\Jcrit {J_\mathrm{crit}}
\newcommand{\rsun}{R_{\odot}}
\newcommand{\mbh}{M_{\rm BH}}
\newcommand{\Msunyr}{M_\odot~{\rm yr}^{-1}}
\newcommand{\mdot}{\dot{M}_*}
\newcommand{\ledd}{L_{\rm Edd}}
\newcommand{\cmc}{{\rm cm}^{-3}}
\def\gsim{~\rlap{$>$}{\lower 1.0ex\hbox{$\sim$}}}
\def\lsim{~\rlap{$<$}{\lower 1.0ex\hbox{$\sim$}}}
\def\Msun {\,\mathrm{M}_\odot}
\def\Jcrit {J_\mathrm{crit}}

\def\simgreat{\lower2pt\hbox{$\buildrel {\scriptstyle >}
   \over {\scriptstyle\sim}$}}
\def\simless{\lower2pt\hbox{$\buildrel {\scriptstyle <}
   \over {\scriptstyle\sim}$}}
\def\msobh{M_\bullet^{\rm sBH}}
\def\zodot{\,{\rm Z}_\odot}
\newcommand{\lambdabar}{\mbox{\makebox[-0.5ex][l]{$\lambda$} \raisebox{0.7ex}[0pt][0pt]{--}}}

\def\na{NewA}%
\def\aj{AJ}%
\def\araa{ARA\&A}%
\def\apj{ApJ}%
\def\apjl{ApJ}%
\def\jcap{JCAP}

\def\pasa{PASA}

\def\apjs{ApJS}%
\def\ao{Appl.~Opt.}%
\def\apss{Ap\&SS}%
\def\aap{A\&A}%
\def\aapr{A\&A~Rev.}%
\def\aaps{A\&AS}%
\def\azh{AZh}%
\def\baas{BAAS}%
\def\jrasc{JRASC}%
\def\memras{MmRAS}%
\def\mnras{MNRAS}%
\def\pra{Phys.~Rev.~A}%
\def\prb{Phys.~Rev.~B}%
\def\prc{Phys.~Rev.~C}%
\def\prd{Phys.~Rev.~D}%
\def\pre{Phys.~Rev.~E}%
\def\prl{Phys.~Rev.~Lett.}%
\def\pasp{PASP}%
\def\pasj{PASJ}%
\def\qjras{QJRAS}%
\def\skytel{S\&T}%
\def\solphys{Sol.~Phys.}%
\def\sovast{Soviet~Ast.}%
\def\ssr{Space~Sci.~Rev.}%
\def\zap{ZAp}%
\def\nat{Nature}%
\def\iaucirc{IAU~Circ.}%
\def\aplett{Astrophys.~Lett.}%
\def\apspr{Astrophys.~Space~Phys.~Res.}%
\def\bain{Bull.~Astron.~Inst.~Netherlands}%
\def\fcp{Fund.~Cosmic~Phys.}%
\def\gca{Geochim.~Cosmochim.~Acta}%
\def\grl{Geophys.~Res.~Lett.}%
\def\jcp{J.~Chem.~Phys.}%
\def\jgr{J.~Geophys.~Res.}%
\def\jqsrt{J.~Quant.~Spec.~Radiat.~Transf.}%
\def\memsai{Mem.~Soc.~Astron.~Italiana}%
\def\nphysa{Nucl.~Phys.~A}%

\def\physrep{Phys.~Rep.}%
\def\physscr{Phys.~Scr}%
\def\planss{Planet.~Space~Sci.}%
\def\procspie{Proc.~SPIE}%

\newcommand{\rmp}{Rev. Mod. Phys.}
\newcommand{\ijmpd}{Int. J. Mod. Phys. D}
\newcommand{\sovjetp}{Soviet J. Exp. Theor. Phys.}
\newcommand{\jkas}{J. Korean. Ast. Soc.}
\newcommand{\PPVI}{Protostars and Planets VI}
\newcommand{\njp}{New J. Phys.}
\newcommand{\rap}{Res. Astro. Astrophys.}

\title{Formation of the First Black Holes}

\setcounter{chapter}{0}

\chapter[Astrophysical black holes]{Astrophysical black holes}\footnotetext{\tiny Preprint of the chapter ``Astrophysical black holes'' of the review volume ``Formation of the First Black Holes'', 2019, Latif, M. and Schleicher, D. R. G., eds., \textcopyright Copyright World Scientific Publishing Company, www.worldscientific.com/worldscibooks/10.1142/10652\par}\label{ch1}

\author[Pedro~R.~Capelo]{Pedro~R.~Capelo}

\address{Center for Theoretical Astrophysics and Cosmology\\Institute for Computational Science\\University of Zurich\\
Winterthurerstrasse 190, CH-8057 Z{\"u}rich, Switzerland\\
pcapelo@physik.uzh.ch}

\begin{abstract}
In this chapter, we introduce the concept of a black hole (BH) and recount the initial theoretical predictions. We then review the possible types of BHs in nature, from primordial, to stellar-mass, to supermassive BHs. Finally, we focus on the latter category and on their intricate relation with their host galaxies.
\end{abstract}

\setcounter{page}{1}

\body

\section{The concept of a black hole and the first predictions}\label{ch1:sec:Black_holes}

All objects in the Universe are gravitational ``holes''. Given the mass and size of a system, there is always a finite speed that needs to be overcome in order to escape its gravitational well. This is the well-known Newtonian concept of the escape velocity, which can be derived by equating the sum of the kinetic and gravitational potential energy of a particle at the surface of the object to zero, and applies to all matter --- from apples to planets, from stars to galaxies. Such a simple calculation was performed by \citet{Michell_1784} --- see also \citet{Laplace_1796} --- when considering light particles with a finite mass and speed, to speculate on the existence of objects (so-called ``dark stars'') from which light could not escape. These objects must therefore have a radius smaller than

\begin{equation}\label{ch1:eq:schwarzschild_radius}
r_{\rm Schw} = \frac{2GM_{\bullet}}{c^2},
\end{equation}

\noindent where $M_{\bullet}$ is the mass of the object, $G$ is the gravitational constant, $c$ is the speed of light in vacuum, and the subscript ``Schw'' refers to Schwarzschild, for reasons that will become apparent below.

The key feature of these objects is not their mass nor their size, but their compactness $C$, defined as the ratio between the value given by Equation~\eqref{ch1:eq:schwarzschild_radius} and the actual size of the system. Planets, stars, and galaxies have values of $C$ of the order of $10^{-9}$--$10^{-6}$, whereas the known stellar compact objects --- white dwarfs and neutron stars --- have $C \sim 10^{-4}$--$10^{-1}$. Only when $C$ reaches 1 do we have a system in which gravity is strong enough that nothing can escape (assuming that $c$ cannot be overcome).

Michell's remarkable intellectual achievement, made using simple Newtonian physics (and assuming the corpuscular theory of light), was confirmed a little more than a century later, shortly after the publication of the theory of general relativity \citep[GR;][]{Einstein_1915,Einstein_1916}, when the first exact vacuum solution to Einstein's field equations was found by \citet{Schwarzschild_1916} --- see also \citet{Droste_1917,Weyl_1917} --- for the case of a spherical, non-electrically charged, non-rotating system. The Schwarzschild metric can be described by \citep[][]{Hilbert_1917}

\begin{equation}\label{ch1:eq:schwarzschild_metric}
{\rm d}s^2 = - \left(1-\frac{r_{\rm Schw}}{r}\right)c^2{\rm d}t^2 + \left(1-\frac{r_{\rm Schw}}{r}\right)^{-1}{\rm d}r^2 + r^2({\rm d}\theta^2 + {\rm d}\phi^2\sin^2\phi),
\end{equation}

\noindent \\where ${\rm d}s^2$ is the space-time line element and the solution is written in spherical coordinates ($t$, $r$, $\theta$, and $\phi$), using the Landau--Lifshitz spacelike convention \citep[][]{Landau_Lifshitz_1962,Misner_et_al_1973}.

Even though the right-hand side of Equation~\eqref{ch1:eq:schwarzschild_metric} diverges at both $r = 0$ and $r_{\rm Schw}$, only the former is a true physical singularity (i.e. the Riemann curvature tensor is infinite only at $r = 0$), with the space-time being non-singular at the so-called Schwarzschild radius.\footnote{This can be easily proven by transforming the coordinates to, e.g. the Kruskal--Szekeres \citep[][]{Kruskal_1960,Szekeres_1960} coordinates.} However, the Schwarzschild radius is of fundamental importance, as the radial coordinate of a particle travelling towards the centre changes from spacelike to timelike when crossing $r_{\rm Schw}$, meaning that the only possible future of that particle is the singularity.\footnote{This is the main difference between the GR solution and Michell's Newtonian dark stars: dark stars can in principle be stable systems, whereas the GR solution {\it must} collapse towards $r = 0$.} Meanwhile, an external static observer will never observe such a boundary (or event horizon) crossing, as the observed time will be infinite (even though the proper time of the particle is finite). Moreover, any radiation sent from such particle and reaching any external observer will be infinitely redshifted. In other words, a photon sent from $r_{\rm Schw}$ would need infinite energy to reach the observer, effectively making the space-time region within the event horizon causally disconnected from the rest of the Universe. For these reasons, objects with an event horizon (i.e. with $C = 1$) are called BHs.

After the publication of the Schwarzschild solution, other exact solutions to Einstein's field equations were found, in the case of electrically charged \citep[][]{Reissner_1916,Nordstrom_1918}, rotating \citep[][]{Kerr_1963}, and rotating, electrically charged BHs \citep[][]{Newman_et_al_1965}. One peculiarity of BHs is that they are extremely simple. In fact, they can be described at most by three parameters: mass, spin (i.e. angular momentum), and electric charge --- this is referred to as the `no hair' theorem \citep[see, e.g.][]{Israel_1967}. Moreover, in typical astrophysical environments, it is believed that electrically charged BHs cannot exist, as any existing electric charge would be quickly cancelled by the charges in the surrounding plasma \citep[or by spontaneous production of pairs of oppositely charged particles; see, e.g.][]{Gibbons_1975,Blandford77}. For this reason, the most complete description for an astrophysical BH is the Kerr metric, which depends only on mass and spin.

Using the Boyer--Lindquist \citep[][]{Boyer_Lindquist_1967} coordinates, one can write down the Kerr metric such that the radial coordinate of the event horizon\footnote{In the Kerr metric, there are actually two event horizons (internal and external). However, the external event horizon described here is, for all intents and purposes, the event horizon of the BH, as it is the first one-way membrane an external particle crosses.} is

\begin{equation}\label{ch1:eq:kerr_event_horizon}
r_{\rm Kerr} = \frac{r_{\rm Schw}}{2}\left(1 + \sqrt{1-a^2}\right),
\end{equation}

\noindent \\where $a \equiv c\mathcal{L}/(GM_{\bullet}^2)$ is the BH spin and $\mathcal{L}$ its angular momentum. The value of $|a|$ can vary between 0 (recovering the Schwarzschild BH) and 1 (for a maximally spinning BH\footnote{The actual maximum value depends on the nature and geometry of the accreting matter \citep[e.g.][]{Bardeen_1970,Thorne_1974,Popham_Gammie_1998}. If $|a| > 1$, the BH would have no event horizon [see Equation~\eqref{ch1:eq:kerr_event_horizon}] and the singularity would be visible by the rest of the Universe, i.e. `naked'. The `cosmic censorship' conjecture \citep[e.g.][]{Penrose_1969} prohibits such cases (similar constraints exist also for electrically charged BHs).}), and its sign depends on the particle orbit we consider: $1$ and $-1$ for corotating and counterrotating orbits (with respect to the BH angular momentum), respectively. If we take a BH with $|a| = 1$, the radial coordinate of the event horizon is half of that of a Schwarzschild BH (i.e. $r_{\rm Kerr} = r_{\rm Schw}/2$).

More importantly, the orbits of massive particles around Kerr BHs vary depending on the value of $a$. When $a = -1$, 0, and 1, the innermost stable circular orbit (ISCO) a massive particle can have is of a radius $r_{\rm ISCO}=$ 4.5, 3, and 0.5 $\times$~$r_{\rm Schw}$, respectively. For $r < r_{\rm ISCO}$, a particle can only spiral inwards (or outwards, if it has enough velocity to do so) and cannot maintain a stable circular orbit.\footnote{We note that, even when $a = 1$, $r_{\rm ISCO}$ is always greater than $r_{\rm Kerr}$ \citep[][]{Bardeen_et_al_1972}.}

The position of the ISCO has significant consequences on how much gravitational energy can be extracted from the gas in the vicinity of the BH \citep[via accretion processes; e.g.][]{Shakura,Blandford_Payne_1982}, as the energy lost by particles increases as the distance from the BH decreases.\footnote{Additionally, for high values of $a$, the ISCO finds itself within the ergosphere, a region of space-time in which, due to the frame-dragging effect \citep[also known as the Lense--Thirring effect;][]{Lense_Thirring_1918}, rotation is inevitable, and this BH rotational energy can be also extracted \citep[e.g.][]{Penrose_1969,Blandford77}.}

In other words, the radiative efficiency $\epsilon_{\rm r}$ (i.e. how much of the rest energy of the accreting particle is released), given by\\

\begin{equation}\label{ch1:eq:epsilon_r}
L = \epsilon_{\rm r} \dot{M}_{\rm \bullet \,accr}c^2,
\end{equation}

\noindent \\\\where $L$ and $\dot{M}_{\rm \bullet \,accr}$ are the accretion power (or luminosity) and the mass accretion rate, respectively, depends on the BH spin \citep[e.g. in a Novikov--Thorne disc, $0.06 \lesssim \epsilon_{\rm r} \lesssim 0.42$ for $0 \le a \le 1$;][]{Novikov_Thorne_1973}.

Indeed, astrophysical accreting BHs are believed to be normally surrounded by accretion discs \citep[e.g.][]{Shakura}, in which matter differentially rotates around the BH, a fraction of it loses angular momentum and moves towards the centre, and, through viscous forces, becomes hot and radiates (e.g. X-rays). The ISCO roughly coincides with the inner edge of the accretion disc. For more on the concept of accretion in astrophysics, see, e.g. \citet{Frank_et_al_2002}.

Accretion on to BHs is of fundamental importance because (a) it is through the results of accretion that we can detect most BHs (e.g. in X-ray binaries and active galactic nuclei --- AGN); (b) it affects the BH nature, \\\\by changing its mass, with the mass growth rate

\begin{equation}\label{ch1:eq:mgrowth}
\dot{M}_{\rm \bullet \,growth} = (1 - \epsilon_{\rm r})\dot{M}_{\rm \bullet \,accr},
\end{equation}

\noindent and spin \citep[e.g.][]{Perego09,Dotti_et_al_2013,Dubois_et_al_2014,Fiacconi_et_al_2017b}; and (c) its power can potentially affect the surrounding material (see Section~\ref{ch1:sec:SMBHs_and_galaxies}). Accretion depends on the availability of low-angular momentum gas\footnote{BHs can also accrete other kinds of matter. For example, BHs with a mass above $\sim$$10^8$~M$_{\odot}$ can ``eat'' whole (solar-mass) stars, without disrupting them. However, in that case the entire mass of the star gets lost inside the BH, and there is no possible extraction of energy (i.e. $\epsilon_{\rm r} = 0$).} in the vicinity of the BH. There is, however, a theoretical limit to how much a BH can accrete, regardless of the availability of ``fuel'', which is given by the radiation pressure --- associated to the accretion power $L$ --- counteracting gravity. When approximating accretion to be spherical \citep[e.g.][]{Hoyle_Lyttleton_1939,Bondi44,Bondi52}, one can equate the inward force on an electron--proton pair --- $GM_{\bullet}m_{\rm p}/r^2$ --- to the outward force --- $L \sigma_{\rm T}/(4 \pi c r^2)$ --- and obtain \citep[][]{Eddington_1916}

\begin{equation}\label{ch1:eq:eddington_luminosity}
L_{\rm Edd} = \frac{4 \pi G M_{\bullet} m_{\rm p} c}{\sigma_{\rm T}},
\end{equation}

\noindent \\where $m_{\rm p}$ is the proton mass and $\sigma_{\rm T}$ is the electron Thomson cross-section. The corresponding mass accretion rate is simply given by $\dot{M}_{\rm \bullet \,Edd} = L_{\rm Edd}/(\epsilon_{\rm r} c^2)$, where $\epsilon_{\rm r}$ is usually chosen equal to 0.1. A BH always accreting at the Eddington level would grow exponentially on a time-scale $\tau_{\rm Salp} = 4.5 \times 10^8\, \epsilon_{\rm r}/(1 - \epsilon_{\rm r})$~yr, known as the Salpeter time \citep[][]{Salpeter_1964}. Assuming, e.g. $\epsilon_{\rm r} = 0.1$, this BH would need $7 \times 10^8$~yr to increase its own mass by six orders of magnitude (e.g. from $10^3$ to $10^9$~M$_{\odot}$). We note that, under certain circumstances (mostly depending on the geometry of the accreting flow), it is still possible to achieve $\dot{M}_{\rm \bullet \,accr} > \dot{M}_{\rm \bullet \,Edd}$ (see Chapter~11).

For a long time after the first GR solution for a BH was found, this was considered only a mathematical curiosity \citep[see, e.g.][]{Einstein_1939}. After all, the size of the Schwarzschild radius is extremely small, ranging from the width of the Manhattan island, when the mass is 1~M$_{\odot}$, to the average distance between the Sun and Uranus, when the mass is $10^9$ times larger. How can so much mass be entrapped in such a small region?

The first answers to this question were given only in the 1930s, with advances in the theory of stellar evolution, when it was postulated that stellar compact objects (e.g. white dwarfs and neutron stars) could indeed exist in nature as the final outcome of the evolution of stars \citep[e.g.][]{Chandrasekhar_1931,Landau_1932,Baade_Zwicky_1934,Tolman_1939,Oppenheimer_Volkoff_1939}, when gravity is not counter-balanced any longer by thermo-nuclear processes. For relatively low-mass stars, gravity is still balanced by some force (electron and neutron degeneracy pressure for white dwarfs and neutron stars, respectively). For higher-mass stars, no force is able to counteract gravity, material flows across the Schwarzschild radius,\footnote{As mentioned above, because of the diverging relation between proper and observer time at $r_{\rm Schw}$, matter infalling through the event horizon seems frozen in time to an observer at infinity, hence the alternative definition of `frozen star' (especially in the Soviet literature).} and a BH forms \citep[][]{Oppenheimer_Snyder_1939}. For a review on the relation between the progenitor's stellar mass and its final fate, see, e.g. \citet{Heger2003,Heger_et_al_2003b}.

For the first time, it was theorised how BHs could indeed form in nature.

\section{Black holes in nature}\label{ch1:sec:Black_Holes_in_Nature}

\begin{figure*}[!t]
\centering
\vspace{-0.0pt}
\includegraphics[width=0.96\columnwidth,angle=0]{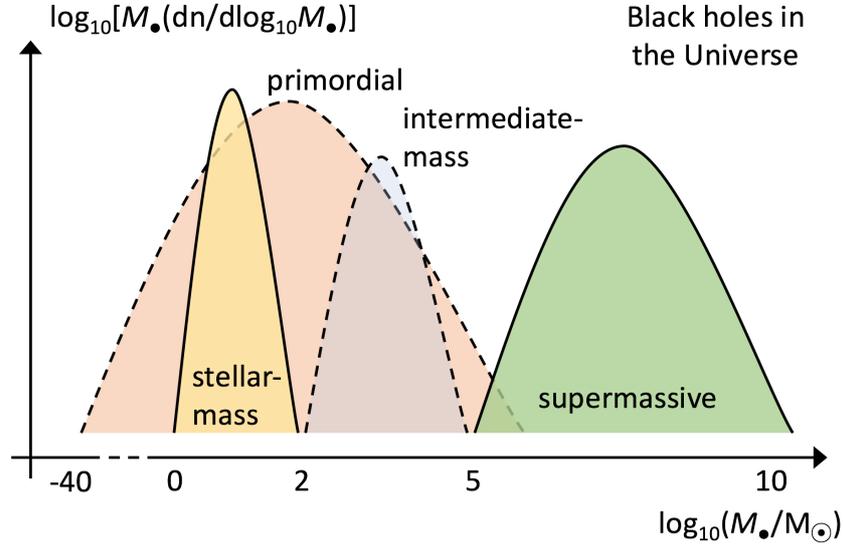}
\vspace{-0.0pt}
\caption[]{Sketched number density --- $M_{\bullet}({\rm d}n/{\rm d}\log_{10} M_{\bullet})$ --- of BHs in the Universe, in the order of lower-mass limit: (from left to right) PBHs, stellar-mass BHs, IMBHs, and SMBHs. The shapes of the distributions of PBHs and IMBHs (dashed lines) are arbitrary, as a reliable census does not exist. In particular, PBHs are highly speculative (see Section~\ref{ch1:sec:Primordial_black_holes}) and IMBHs have not been detected yet (although there are a few strong candidates; see Section~\ref{ch1:sec:Intermediate_mass_black_holes}). The small number of detected stellar-mass BHs does not allow for an accurate measurement of their mass function, but see, e.g. \citet{Merloni_2008,Kovetz_et_al_2017}. The mass function of SMBHs is inspired by \citet{Shankar_et_al_2009}. Figure inspired by \citet{Garcia-Bellido_2017}.
}
\vspace{-4.0pt}
\label{ch1:fig:BH_mass_distribution}
\end{figure*}

A BH can in principle have any mass, as long as there is a mechanism able to compress such mass inside its own event horizon. In this section, we divide the BHs believed to exist in nature in the order of their lower-mass limit, starting from the highly speculative primordial BHs (PBHs; Section~\ref{ch1:sec:Primordial_black_holes}), whose masses could range from $\sim$$10^{-5}$~g to $\sim$$10^5$~M$_{\odot}$ (or more), continuing with stellar-mass BHs ($\sim$3--100~M$_{\odot}$; Section~\ref{ch1:sec:Stellar_mass_black_holes}), and ending with intermediate-mass BHs (IMBHs; $\sim$100--$10^5$~M$_{\odot}$; Section~\ref{ch1:sec:Intermediate_mass_black_holes}) and supermassive BHs (SMBHs;  $\gtrsim$$10^5$~M$_{\odot}$; Section~\ref{ch1:sec:Supermassive_black_holes}).

There are several methods for measuring the mass of a BH, ranging from dynamical measurements of individual stars (e.g. X-ray binaries and Sagittarius A$^*$; see Sections~\ref{ch1:sec:Stellar_mass_black_holes} and \ref{ch1:sec:Supermassive_black_holes}) and gas clouds (e.g. through water masers; NGC~4258; see Section~\ref{ch1:sec:Supermassive_black_holes}), ensembles of stars (from stellar absorption lines), and gas (from gas emission lines in, e.g. circumnuclear discs), to less direct methods such as reverberation (or echo) mapping \citep[e.g.][]{Blandford_McKee_1982,Peterson_1993}, simple Eddington-luminosity arguments, or even using virial estimates from known scaling relations (see Section~\ref{ch1:sec:SMBHs_and_galaxies}). The technique used mostly depends on whether the BH is accreting and/or if we can resolve its sphere of influence, defined by the radius\footnote{An alternative definition is given by $M_{\star} (r < r_{\rm infl})  = 2 M_{\bullet}$, where $M_{\star}$ is the stellar mass. The two definitions coincide if the galaxy can be described by a singular isothermal sphere \citep[e.g.][]{Merritt_2013}. A related (and of the same order of magnitude) quantity is the Bondi radius \citep[][]{Bondi52}, given by $G M_{\bullet}/c_{\rm s}^2$, where $c_{\rm s}$ is the sound speed of the surrounding gas, within which the gas is gravitationally bound to the BH.}

\begin{equation}\label{ch1:eq:sphere_of_influence}
r_{\rm infl} = \frac{GM_{\bullet}}{\sigma^2},
\end{equation}

\noindent where $\sigma$ is the stellar velocity dispersion of the central part of the galaxy (assumed to be constant with radius), within which the BH gravitational potential dominates the motion of the surrounding matter. E.g. a BH of mass $10^8$~M$_{\odot}$ in a galaxy with $\sigma = 200$~km~s$^{-1}$ has a sphere of influence of radius $\sim$10~pc.

In Figure~\ref{ch1:fig:BH_mass_distribution}, we show a rough sketch of the mass distribution of BHs in the Universe.

\subsection{Primordial black holes}\label{ch1:sec:Primordial_black_holes}

According to the most accepted theory,\footnote{Amongst the alternative theories, we list, e.g. cosmic string loops \citep[e.g.][]{Hawking_1989} and bubble collisions \citep[e.g.][]{Hawking_et_al_1982}.} PBHs are believed to be the by-product of the extremely high densities and inhomogeneities present in the first instants after the Big Bang \citep[][]{Zeldovich66,Hawking_1971}, when the Universe was radiation-dominated. The PBH mass is presumed to be of the order of the particle horizon mass \citep[e.g.][]{Carr_2005}:

\begin{equation}\label{ch1:eq:PBH_mass}
M_{\bullet} \sim \frac{c^3t_{\rm form}}{G} \simeq 4 \times 10^{38} \frac{t_{\rm form}}{\rm s}~{\rm g} \simeq 2 \times 10^5 \frac{t_{\rm form}}{\rm s}~{\rm M}_{\odot},
\end{equation}

\indent \\where $t_{\rm form}$ is the moment when they formed. Depending on when the PBH forms, its mass can then span an enormous range of values, from the Planck mass (when $t_{\rm form} = t_{\rm Planck}$) to $10^5$~M$_{\odot}$ (when $t_{\rm form} \sim 1$~s) or larger.

A particularly important value is $10^{15}$~g (when $t_{\rm form} \sim 10^{-23}$~s), because it is approximately the mass of PBHs which should be evaporating today due to the \citet{Hawking_1974} effect,\footnote{According to the Hawking effect, a particle--antiparticle pair can be produced from the BH's gravitational energy in the vicinity of the event horizon. If, instead of immediately recombining, one of the particles falls into the BH and the other escapes, the net effect is that the BH has lost some of its energy (i.e. its mass), corresponding to the energy of the escaping particle.} since BHs should evaporate with a time-scale $\sim 10^{71} (M_{\bullet}/{\rm M}_{\odot})^3~$s. Larger BHs will evaporate in the future, whereas smaller BHs may have already evaporated.\footnote{We note, in passing, that any BH that may be created in the laboratory (e.g. in the Large Hadron Collider), would be so small that it would evaporate almost instantly \citep[e.g.][]{Dimopoulos_Landsberg_2001}.} So far, however, no firm detection of Hawking radiation has been reported in the extragalactic or Galactic $\gamma$-ray backgrounds \citep[e.g.][]{Carr_et_al_2016a}.

PBHs with a mass larger than $10^{15}$~g should in principle be still present today, and they have also been touted as partly or fully responsible for the dark matter in the Universe. The recent detections of gravitational waves (GWs) from the advanced Laser Interferometer Gravitational-wave Observatory (LIGO) of the inspiral and coalescence of BHs of masses somewhat larger than usually expected from stellar evolution (e.g. \citealt{Abbott2016A}; see also Section~\ref{ch1:sec:Stellar_mass_black_holes} and Figure~\ref{ch1:fig:Xray_binaries}) have given new life to this scenario, which dates back to the MACHO --- massive, compact, halo object --- era \citep[e.g.][]{Paczynski_1986,Alcock2000}. To date, however, there are several gravitational-lensing and dynamical constraints that likely make such a scenario implausible \citep[for a recent review, see, e.g.][]{Carr_et_al_2016b}.

Finally, PBHs could actually be the seeds of the SMBHs found at the centre of massive galaxies \citep[see, e.g.][]{Bernal_et_al_2018}.

\subsection{Stellar-mass black holes}\label{ch1:sec:Stellar_mass_black_holes}

\begin{figure*}[!t]
\centering
\vspace{-0.0pt}
\hspace{48.0mm}\begin{overpic}[width=0.96\columnwidth,angle=0,trim={-2cm 0 0 0},clip]{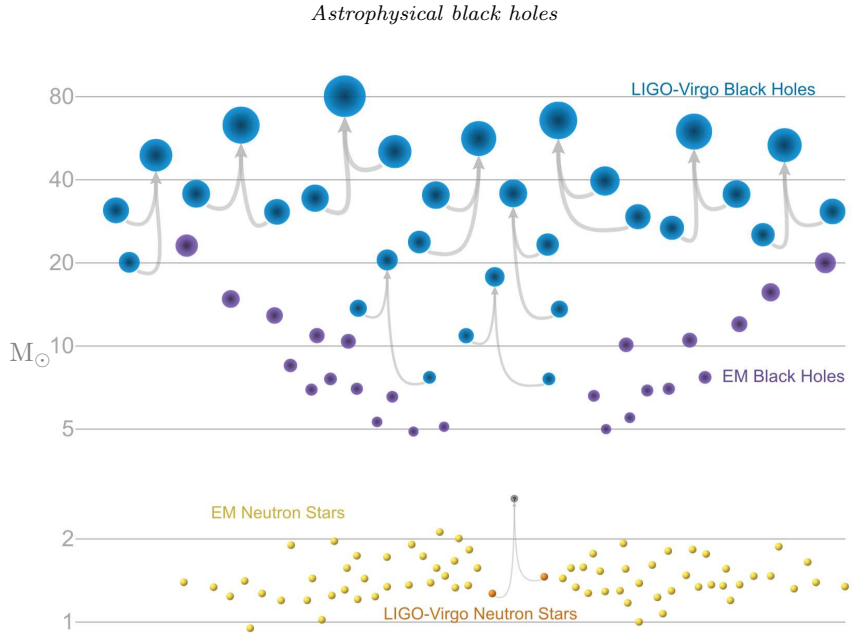} 
\put (-1,33) {\textcolor{gray}{M$_{\odot}$}}
\end{overpic}
\vspace{-0.0pt}
\caption[]{Masses of stellar-mass BHs (blue and purple; mass $\gtrsim 3$~M$_{\odot}$) and neutron stars (orange and yellow; mass $\lesssim 3$~M$_{\odot}$) detected through electromagnetic (EM; purple and yellow) and GW observations (blue and orange). The heaviest stellar-mass BHs in the plot have masses in the range $\sim$60--80~M$_{\odot}$ \citep[GW150914, GW170823, and GW170729;][]{Abbott2016A,GWTC1}. The nature of the remnant of the LIGO-Virgo neutron-star system \citep[GW170817;][]{Abbott_et_al_2017}, marked by a question mark (of mass $\sim$3~M$_{\odot}$), is still unknown. Image credit: LIGO-Virgo/Frank Elavsky/Northwestern University.
}
\label{ch1:fig:Xray_binaries}
\end{figure*}

Stellar-mass BHs have masses of the order of $\sim$3--100~M$_{\odot}$ and are believed to be the final outcome of the evolution of massive stars (see Section~\ref{ch1:sec:Black_holes}). Given their size and distance, it is currently (and for the foreseeable future) extremely difficult to directly detect such systems in isolation.

Luckily, most massive stars are in binaries and, as they coevolve, the end state of one may be a compact object, which accretes mass from the other that is in its late evolutionary stages and is blowing off its outer envelope. Hence the presence of a BH (or neutron star) is inferred from evidence for episodic accretion, inferred from the light curves and spectral energy distributions of variable sources.

\citet{Giacconi_et_al_1962}, in the first X-ray exploration of space, discovered the source Cygnus X-1, which was later confirmed to be a binary system composed of a stellar-mass BH \citep[of mass $\sim$15~M$_{\odot}$;][]{Orosz_et_al_2011} and a companion star. Nowadays, several X-ray binaries have been detected (see Figure~\ref{ch1:fig:Xray_binaries}), in which the X-ray radiation is believed to be originated from the companion-star gas being accreted on to the compact object.

In many cases, it is assumed that the compact object is a neutron star. In some cases, however, as in the case of Cygnus X-1, it is highly probable that the compact object is a BH, since the calculated mass is very likely above the upper limit for neutron stars \citep[][]{Tolman_1939,Oppenheimer_Volkoff_1939}. The uncertainty arises from the fact that we cannot measure directly the mass, but only the mass function, defined as

\begin{equation}\label{ch1:eq:mass_function}
f(M_{\bullet}) \equiv \frac{M_{\bullet} \sin^3 i}{(1+q)^2} = \frac{Pv^3_{\rm r}}{2 \pi G},
\end{equation}

\noindent \\where $P$ is the orbital period, $q$ is the mass ratio, $v_{\rm r}$ is the radial velocity of the companion star, and $i$ is the orbital inclination. Radial-velocity measurements provide only a lower limit on the mass, mostly due to the difficulty of determining the angle $i$ (unless the binary is eclipsing).

When a stellar-mass compact object is not accreting gas, it is virtually impossible to detect such an object.\footnote{However, microlensing surveys such as MACHO \citep[][]{Bennett_et_al_2002} have provided a few candidates.} Only very recently, thanks to the first detections of GWs by the advanced LIGO, it has been possible to detect pairs of stellar-mass compact systems (both BHs and neutron stars) in the final stages of a merger.\footnote{With the planned Laser Interferometer Space Antenna (LISA), it will also be possible to detect extreme mass ratio inspiral (EMRI) events between a stellar-mass BH and a SMBH \citep[e.g.][]{Babak_et_al_2017}.} By comparing the detection from interferometers to numerical-GR models, both the mass of the merging systems and that of the remnant were inferred, yielding a LIGO-BH mass range that was slightly on the upper end of what usually thought possible from stellar-evolution modelling (\citealt{Heger2003,Heger_et_al_2003b}; see also Section~\ref{ch1:sec:Primordial_black_holes}). GW detections have provided a new way to study mergers of compact objects. For example, stellar-mass compact-object mergers (BH--neutron star or neutron star--neutron star) are considered an important pathway for the production of the elements beyond iron group nuclei \citep[$r$-process nucleosynthesis; see, e.g.][]{Lattimer_Schramm_1974,Rosswog_2015,Bonetti_et_al_2018}. Very recently, the first neutron star--neutron star merger was detected \citep[GW170817;][]{Abbott_et_al_2017}. Hopefully, we will soon be able to detect also BH--neutron star coalescences.\vspace{-6.0pt}

\subsection{Intermediate-mass black holes}\label{ch1:sec:Intermediate_mass_black_holes}

IMBHs have a mass range of $\sim$100--$10^5$~M$_{\odot}$ and are, amongst the non-primordial BHs, the most elusive of all.

At high redshift $z$, they are believed to be one of the types of BH seeds that will eventually become SMBHs. Given their relatively low mass, it is currently extremely difficult to detect them and measure their mass both by gas or stellar dynamics methods (as their sphere of influence can be smaller than 1~pc, depending on the host) or by emission from accreting gas [as the Eddington luminosity scales linearly with $M_{\bullet}$; see Equation~\eqref{ch1:eq:eddington_luminosity}].

In this section, we bring our attention to those seed BHs which, for several reasons, did not grow to become SMBHs and should therefore have masses in the IMBH range today.

One obvious place to look for IMBHs are local dwarf galaxies: these galaxies are thought to have had a relatively quiet life, with no major mergers nor significant gas accretion, hence they are not too dissimilar from their high-redshift counterparts. It is plausible, therefore, that an existing central BH has not accreted much during its life and has retained its initial relatively low mass. Detections of such BHs in local dwarf galaxies would be very important also to understand how they formed in the first place, since different formation scenarios would produce different BH occupation fractions and scaling relations in the low-mass range \citep[e.g.][]{Volonteri10}. Several IMBH candidates have been identified in dwarf galaxies \citep[e.g. RGG~118; $\sim$$5 \times 10^4$~M$_{\odot}$; ][]{Baldassare15} and populations of IMBHs in dwarf galaxies have been recently detected up to $z \sim 2$ \citep[e.g.][]{Mezcua_et_al_2018}.

Another site where IMBHs could hide are globular clusters, since BHs could form from runaway collapse of stellar clusters (see Chapter~7). However, so far, all BH-mass estimates from kinematic measurements were shown to be also explained without a BH \citep[e.g. NGC~7078;][]{vandenBosch_et_al_2006}. Moreover, strong evidence for accretion has not been found, as globular clusters are likely devoid of gas. Recently, \citet{Kiziltan_et_al_2017} used a novel method, by measuring the locations and accelerations of pulsars in the globular cluster 47 Tucanae (NGC~104), comparing them to what was modelled assuming no BH, and computing the mass of the central BH necessary to match the observations, obtaining a mass of $\sim$$2.3 \times 10^3$~M$_{\odot}$.

IMBHs have also been linked to ultra-luminous X-ray sources (ULXs), which are non-nuclear, point-like extragalactic sources with an apparent isotropic luminosity $>$$10^{39}$~erg~s$^{-1}$ in the 0.3--10~keV band \citep[e.g.][]{Feng11}. Such luminosities require either a relatively massive BH accreting at Eddington levels [see Equation~\eqref{ch1:eq:eddington_luminosity}] or stellar-mass compact objects accreting at super-Eddington levels. The growing consensus is that most ULXs are indeed super-Eddington accreting stellar-mass compact objects \citep[either BHs or neutron stars; e.g.][]{Bachetti_et_al_2014,Fiacconi_et_al_2017a} with only the highest-luminosity ULXs (hyper-luminous X-ray sources --- HLXs; $\gtrsim$$5 \times 10^{40}$~erg~s$^{-1}$) possibly related to IMBHs \citep[e.g. HLX-1, with a mass range $\sim$$6 \times 10^3$--$2 \times 10^5$~M$_{\odot}$, depending on inclination and BH spin;][]{Farrell_et_al_2009,Straub_et_al_2014}.

IMBHs could be found also in other ways and environments: within accretion discs of SMBHs \citep[e.g.][]{Bellovary_et_al_2016}; hidden inside high-velocity clouds near SMBHs \citep[e.g.][]{Oka_et_al_2016}; or detected by future gravitational-wave interferometers \citep[e.g. LISA;][]{Miller_2009,Tamfal_et_al_2018} when two of them coalesce. Given their relatively small mass, they are also able to tidally disrupt (solar-mass) stars \citep[e.g.][]{Hills_1975,Rees_1988} --- as opposed to eating them whole, as is the case for SMBHs with mass $\gtrsim 10^8$~M$_{\odot}$ --- possibly producing an ultraviolet or X-ray signal when the debris gets accreted \citep[e.g.][]{Evans_Kochanek_1989,Renzini_et_al_1995,Bade_et_al_1996,Bloom_et_al_2011,Mainetti_et_al_2016}.

For a recent review on IMBHs and their observational evidence, see, e.g. \citet{Mezcua_2017}.

\subsection{Supermassive black holes}\label{ch1:sec:Supermassive_black_holes}

\begin{figure*}[!t]
\centering
\vspace{-0.0pt}
\includegraphics[width=0.400\columnwidth,angle=0]{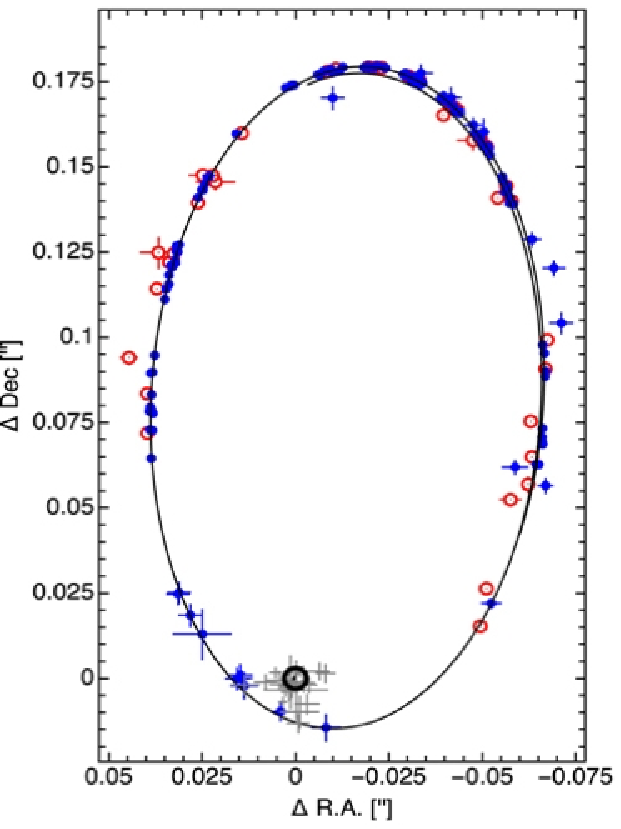}
\includegraphics[width=0.575\columnwidth,angle=0]{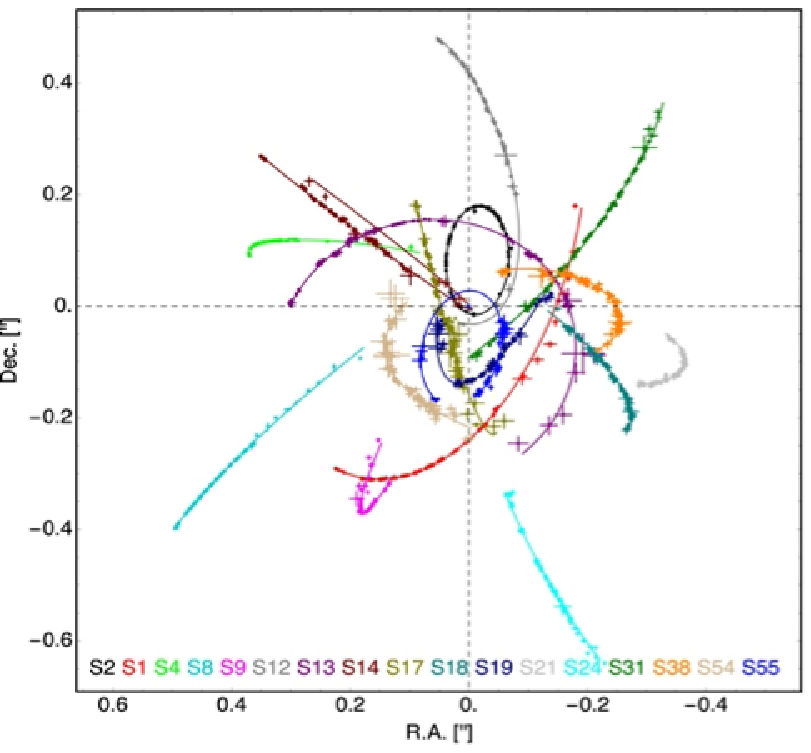}
\vspace{-8.0pt}
\caption[]{Orbit of the star S2 (left-hand panel) and of 17 S-stars (right-hand panel) around Sagittarius A$^*$. Figures from \citet{Gillessen_et_al_2017}.  \textcopyright AAS. Reproduced with permission.}
\vspace{-10.0pt}
\label{ch1:fig:SgrAstar}
\end{figure*}

SMBHs have masses greater than $\sim$$10^5$~M$_{\odot}$ and can reach values of a few $10^{10}$~M$_{\odot}$. In principle, there should be no upper limit to the mass, but both observations and theoretical reasoning \citep[see, e.g.][]{Natarajan_Treister_2009,King_2016,Inayoshi_Haiman_2016,Pacucci_et_al_2017} seem to imply that SMBHs heavier than a few $10^{10}$~M$_{\odot}$ do not exist \citep[or, if they do, they grow their mass above the upper limit by non-luminous means such as SMBH mergers;][]{King_2016}.

The first hint that extremely large BHs could exist in nature came with the first observations of quasars in the early 1960s.\footnote{Quasars, or quasi-stellar radio sources, are systems in which the host galaxy is not resolved and only the central point source's emission is detected. They are a subset of AGN \citep[see reviews by][]{Peterson_1997,Krolik_1998,Ho_2008,Padovani_et_al_2017}. In this chapter, we use the two terms interchangeably.} These systems were extremely distant \citep[e.g.][]{Schmidt_1963}, implying that they had luminosities exceeding even $10^3$ times the typical luminosities of galaxies. In many cases, they had jets whose direction was the same over distances ranging from sub-pc to several tens of kpc. In some of these cases, the jet showed (apparent) superluminal motion, which can be explained by a jet pointing towards us and with velocities very close to $c$ \citep[e.g.][]{Rees_1966}. Also, many sources showed time variability of the order of even weeks.

All these pieces of indirect evidence point to the possibility that quasars are indeed powered by SMBHs \citep[][]{Salpeter_1964,Zeldovich_1964,Lynden-Bell_1969}. Amongst all possible astrophysical objects that we know of, only BHs have such a deep potential well that can produce so much accretion energy. Also, only a relativistic potential can create jets with velocities close to the speed of light, and the spin of a BH is a plausible cause for the stability of a jet over large time- and length-scales. Finally, time variability of the order of weeks implies sizes of the order of `$c$~$\times$~week' and, consequently, high values of $C$.

Since the first observations of quasars, a plethora of systems, all under the umbrella-name of AGN, have been observed at all wavelengths. Additionally, systems observed earlier than the first quasars \citep[e.g. Seyfert galaxies;][]{Seyfert_1943} were also catalogued as AGN. AGN are by definition accreting SMBHs (and IMBHs). Only when there is enough gas fuelling (i.e. enough gas losing angular momentum), a SMBH can shine through its accretion disc as an AGN and be easily detected.

When SMBHs are not accreting, their existence can still be inferred (and their masses can still be measured) through dynamical methods. Here we list only the two strongest cases.

The best example is the SMBH at the centre of our own Milky Way, coincident with the radio source Sagittarius A$^*$. Repeated observations of individual stellar proper motions have shown almost perfect Keplerian orbits (see Figure~\ref{ch1:fig:SgrAstar}), implying the existence of a very compact object. In particular, the complete orbit of the S2 (or S0-2) star shows that there is a mass of $\sim$$4.3 \times 10^6$~M$_{\odot}$ within a region of $\sim$100~AU \citep[e.g.][]{Ghez_et_al_2008,Genzel2010}.

Another strong case is in the galaxy NGC~4258, in which water masers (water molecules in gas clouds, excited by the central AGN) were found to trace a well-organised (Keplerian) rotating thin disc, viewed edge-on, whose kinematics could be explained only by a central compact object of mass $\sim$$4 \times 10^7$~M$_{\odot}$ \citep[][]{Miyoshi_et_al_1995}.

\begin{figure*}[!t]
\centering
\vspace{-0.0pt}
\includegraphics[width=0.98\columnwidth,angle=0]{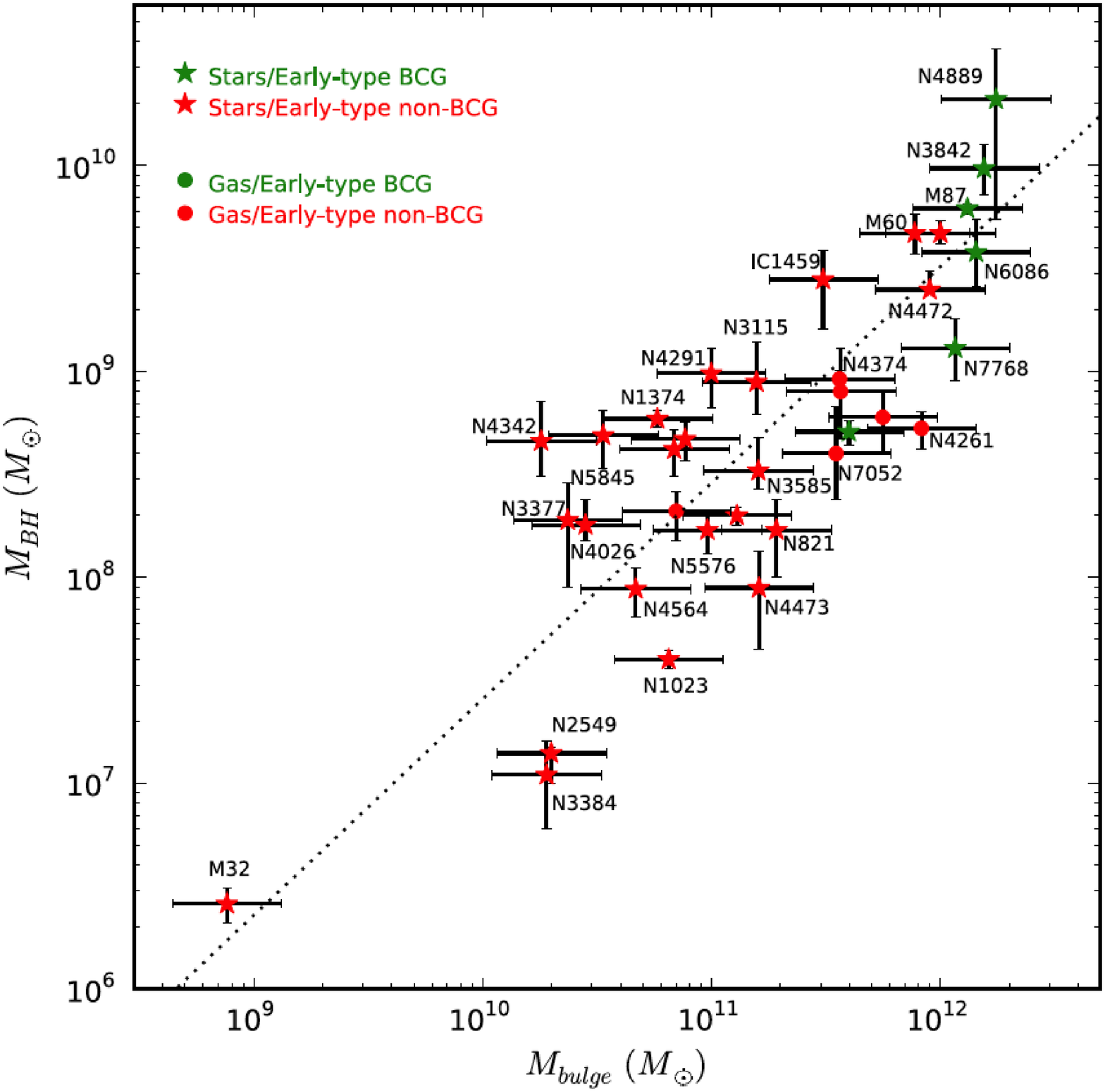}
\vspace{-0.0pt}
\caption[]{Relation between the mass of the central SMBH $M_{\bullet}$ and the bulge mass $M_{\rm bulge}$ of the host galaxy, for the 35 early-type galaxies with dynamical measurements of the bulge stellar mass in the sample of \citet{McConnell_Ma_2013}, divided in brightest cluster galaxies (BCG; green) and non-BCG (red). The SMBH masses are measured using the dynamics of stars (stars) or gas (circles). The error bars indicate 68 per cent confidence intervals. The black, dotted line represents the best-fitting power-law relation, which is in good agreement with Equation~\eqref{ch1:eq:magorrian_relation}. Figure from \citet{McConnell_Ma_2013}.  \textcopyright AAS. Reproduced with permission.}
\label{ch1:fig:Magorrian}
\vspace{2.0pt}
\end{figure*}

In both these cases, it is still in principle possible to imagine alternatives to SMBHs:\footnote{More exotic alternatives include, e.g. fermion stars \citep[currently ruled out;][]{Schodel_et_al_2002}, boson stars, and gravastars.} e.g. a cluster of non-luminous objects (e.g. stellar-mass BHs, neutron stars, white dwarfs, brown dwarfs) could in fact be the dark system at the centre of both the Milky Way and NGC~4258. However, it was shown by \citet{Maoz_1998} that such systems would collapse on a time-scale much shorter than the galaxy's age. Moreover, recent detailed observations of the orbit of the S2 star around the Galactic centre provided the detection of gravitational redshift \citep[][]{GRAVITY}. Very recently, the Event Horizon Telescope, a global very long baseline interferometry array, resolved the shadow of the SMBH in the giant elliptical galaxy M87 (NGC 4486), providing for the very first time a ``picture'' of a BH \citep[][]{EHT}.

One fundamental and still unanswered question is: how did SMBHs form? Several alternatives have been put forward for the origin of the seeds \citep[e.g.][]{Rees_1978,Volonteri12}, including, e.g. direct collapse of pristine gas (\citealt{Haehnelt1993}; see Chapters~5 and 6), population~III BHs (\citealt{Madau-Rees01}; see Chapter~4), dense stellar clusters (\citealt{Begelman1978}; see Chapter~7), or galaxy mergers (\citealt{Mayer2010}; see Chapter~5). SMBH formation is the main topic of this book, on which therefore we will not dwell further in this section. In the remainder of this chapter, we will discuss the importance of SMBHs in relation to their host galaxies.

\section{Supermassive black holes and their host galaxies}\label{ch1:sec:SMBHs_and_galaxies}

It is now widely accepted that SMBHs exist at the centre of several, if not most, nearby massive galaxies \citep[e.g.][]{Kormendy2013}. Additionally, SMBHs as the engines of AGN are believed to exist in many high-redshift galaxies.

In many instances, the mass of the central SMBH has been estimated via a wide variety of methods (see Sections~\ref{ch1:sec:Black_Holes_in_Nature} and \ref{ch1:sec:Supermassive_black_holes}). Accurate mass measurements in the local Universe led to the finding of scaling relations between the mass of the central object and several galactic quantities: stellar velocity dispersion, bulge stellar mass (see Figure~\ref{ch1:fig:Magorrian}), bulge optical and near-infrared luminosity, total luminosity, bulge concentration, dark matter halo mass, number and total mass of globular clusters in the host galaxy, light deficit, spiral-arm pitch angle, etc. \citep[see, e.g.][and references therein]{McConnell_Ma_2013}.

In particular, tight scaling relations have been found between the SMBH mass and a few properties of the host, including the bulge luminosity \citep[e.g.][]{Kormendy_Richstone_1995}, bulge mass $M_{\rm bulge}$ \citep[the so-called Magorrian relation; e.g.][]{Magorrian_et_al_1998},

\begin{equation}\label{ch1:eq:magorrian_relation}
\frac{M_{\bullet}}{10^9 {\rm M}_{\odot}} = 0.49^{+0.06}_{-0.05} \left(\frac{M_{\rm bulge}}{10^{11} {\rm M}_{\odot}}\right)^{1.17 \pm 0.08},
\end{equation}

\noindent and effective velocity dispersion $\sigma$ \citep[the so-called $M_{\bullet}$--$\sigma$ relation; e.g.][]{Ferrarese00,Gebhardt00},

\begin{equation}\label{ch1:eq:Msigma_relation}
\frac{M_{\bullet}}{10^9 {\rm M}_{\odot}} = 0.310^{+0.037}_{-0.033} \left(\frac{\sigma}{200~{\rm km~s}^{-1}}\right)^{4.38 \pm 0.29},
\end{equation}

\noindent \\where the coefficients of the fits of Equations~\eqref{ch1:eq:magorrian_relation} and \eqref{ch1:eq:Msigma_relation} are valid for the classical bulges and elliptical galaxies listed in \citet{Kormendy2013}.

\begin{figure*}[!t]
\centering
\vspace{-0.0pt}
\includegraphics[width=0.95\columnwidth,angle=0]{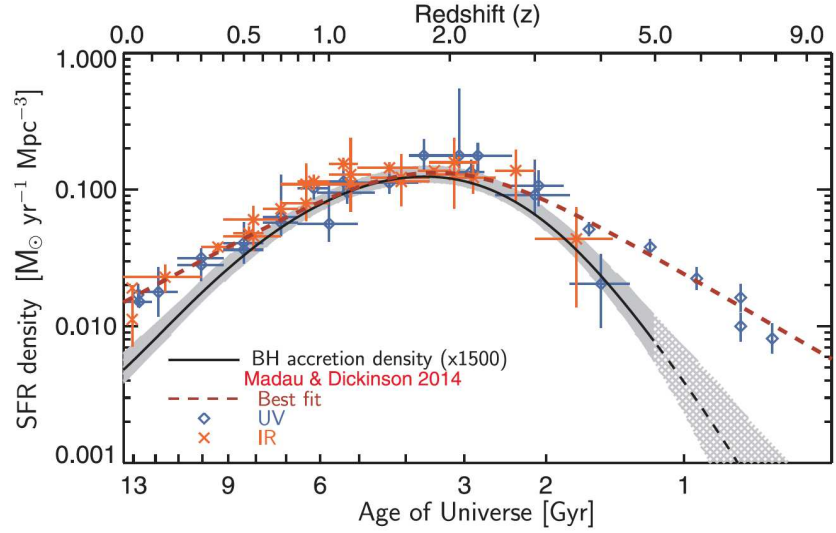}
\vspace{-0.0pt}
\caption[]{SFR density [measurements based on rest-frame ultraviolet (blue diamonds) and infrared (orange crosses) observations, together with their best fit (dark-red, dashed line), from \citet{Madau14-review}] and SMBH accretion rate density [black line and 99 per cent confidence shaded region, from the model of \citet{Aird_et_al_2015}, extrapolated for $z > 5$], scaled up by a factor of $1.5 \times 10^3$, as a function of redshift. See also the second and third panel of Figure~\ref{ch1:fig:capelo}. Figure from \citet{Aird_et_al_2015}, reproduced by permission of Oxford University Press, on behalf of the Royal Astronomical Society.
}
\label{ch1:fig:Madau_plot}
\end{figure*}

Moreover, high-redshift observations have shown that the evolution of the cosmic star formation rate (SFR) and the SMBH accretion rate densities over cosmic time follow very similar trends (e.g. \citealt{Madau14-review,Aird_et_al_2015}; see Figure~\ref{ch1:fig:Madau_plot}): both rise up to $z \sim 2$, where they reach their peak, before steeply declining down to local values. Remarkably, the ratio between the two remains more or less the same across all cosmic time, with the SFR $\sim$$10^3$ times larger than the SMBH accretion rate.

All the above considerations are consistent with a scenario in which SMBHs and their host galaxies may co-evolve\footnote{We caution that there are alternative explanations for the origin of scaling relations, which do not require any co-evolution. E.g. the tightness of the Magorrian relation could be partly or fully caused simply by the hierarchical assembly of SMBH and stellar mass through galaxy merging, as a consequence of the central-limit theorem \citep[e.g.][]{Peng_2007,Hirschmann_et_al_2010,Jahnke_Maccio_2011}.} \citep[e.g.][]{Silk98,Haehnelt_et_al_1998}: SMBHs grow through gas accretion and mergers and feed back part of their accretion energy to the host galaxy.

Assuming that SMBHs grow mostly due to radiative accretion (a reasonable assumption, as we will see below), the energy released by a SMBH to grow its mass to $M_{\bullet}$ is $E_{\bullet} \simeq \epsilon_{\rm r}M_{\bullet}c^2/(1 - \epsilon_{\rm r})$ [see Equations~\eqref{ch1:eq:epsilon_r} and \eqref{ch1:eq:mgrowth}]. The binding energy of the host galaxy bulge is instead $E_{\rm bulge} \simeq M_{\rm bulge} \sigma^2$. Combining Equations~\eqref{ch1:eq:magorrian_relation} and \eqref{ch1:eq:Msigma_relation} and taking $\epsilon_{\rm r} = 0.1$, we obtain $E_{\bullet}/E_{\rm bulge} \simeq 8 \times 10^2 (M_{\bullet}/10^9 {\rm M}_{\odot})^{-0.3}$. The energy released by the growth of a SMBH of mass $10^7$~M$_{\odot}$ ($10^{10}$~M$_{\odot}$) is $\sim$$3 \times 10^3$ ($\sim$$4 \times 10^2$) times the binding energy of the host bulge. If even a small part of $E_{\bullet}$ couples to the gas of the galaxy, all this gas would be blown away \citep[e.g.][]{Fabian12,King_Pounds_2015}. Clearly, there is enough energy in the accretion process to have a significant effect on the host galaxy. This so-called AGN feedback can be radiative and/or mechanical (jets/winds) and has been invoked to explain, e.g. the above-mentioned scaling relations and the mismatch between the dark matter halo mass function and the observed luminosity function (at large masses). Feedback is usually negative (i.e. SF is quenched) and is either ejective (with the gas being blown away from the galaxy) or preventive (with the gas being kept hot), but can also be positive (i.e. SF is enhanced), mostly in the case of jets/winds. The topic of AGN feedback is at the same time still hotly debated and so vast that it is beyond the scope of this introductory chapter. We refer the reader to recent reviews by, e.g. \citet{Alexander_Hickox_2012,Fabian12,Kormendy2013,Heckman_Best_2014,King_Pounds_2015}.

One can check the assumption that SMBHs grow due to radiative accretion, by comparing the AGN activity at high redshift to the mass of local SMBHs. This is the basis of the so-called So\l{}tan argument \citep[][]{Soltan82}. By integrating the AGN (bolometric) luminosity over redshift (thus obtaining the comoving energy density from AGN), one can then obtain the SMBH mass density at $z = 0$ and compare it to local observations (which rely on the above-mentioned scaling relations). The overall agreement between the two numbers \citep[see, e.g.][]{Yu_Tremaine_2002,Marconi04}, when assuming a radiative efficiency $\epsilon_{\rm r} \sim 0.1$, strengthens the hypothesis that SMBHs grow mostly by radiative accretion.\footnote{An individual SMBH can also grow via SMBH mergers, radiatively inefficient accretion \citep[e.g. in advection-dominated accretion flows;][]{Narayan_Yi_1994}, or by swallowing whole stars (see also Section~\ref{ch1:sec:Intermediate_mass_black_holes}), although the local SMBH mass density is barely affected by these mechanisms. We note, in passing, that SMBHs can affect individual stars also in opposite fashion, by disrupting a tight binary stellar system and causing one of the stars to be ejected at velocities of the order of $10^3$~km~s$^{-1}$, hence taking the name of hyper-velocity star \citep[e.g.][]{Hills_1988}.}

If almost all the mass of a SMBH is indeed provided by radiative accretion, this brings us to the question: how can gas get accreted efficiently?

The existence of high-redshift quasars with SMBH masses of the order of $10^9$~M$_{\odot}$ implies that a comparable amount of gas was funnelled into the very inner galactic regions in a very short time: the age of the Universe at $z \sim 7.54$, the redshift of the current record-holder quasar \citep[][]{Banados18}, was $\sim$0.7~Gyr.

The main issue to overcome is not the quantity of available gas, but the amount of angular momentum that such gas needs to lose in order to travel from kpc scales (galactic or even extra-galactic regions) to sub-pc scales (the accretion disc). If we take a typical galaxy with a flat rotation velocity $v = 200$~km~s$^{-1}$, gas at $r = 10$~kpc from the centre has a specific angular momentum magnitude $l = rv \simeq 2 \times 10^{3}$~kpc$^2$~Gyr$^{-1}$. If we now compute the same quantity at the ISCO radius of a maximally spinning Kerr BH ($r = r_{\rm Schw}/2$; see Section~\ref{ch1:sec:Black_holes}), we obtain $l = (GM_{\bullet}r)^{1/2} \simeq 1.5 \times 10^{-11} (M_{\bullet}/{\rm M}_{\odot})$~kpc$^2$~Gyr$^{-1}$. Taking, e.g. a BH of mass $10^9$~M$_{\odot}$, this means that the gas needs to lose 99.999 per cent of its specific angular momentum.

Gas can lose its angular momentum through many different mechanisms, depending on the state of the system (e.g. if it is isolated or interacting) and on the distance from the centre \citep[for a review, see, e.g.][]{Jogee_2006}. Most of these mechanisms are linked to deviations from axial symmetry in the gravitational potential, so that a torque can act on the gas (but can also rely on, e.g. dynamical friction, as individual gas clouds travel inwards).

\begin{figure*}[!t]
\centering
\vspace{-0.0pt}
\includegraphics[width=0.64\columnwidth,angle=0]{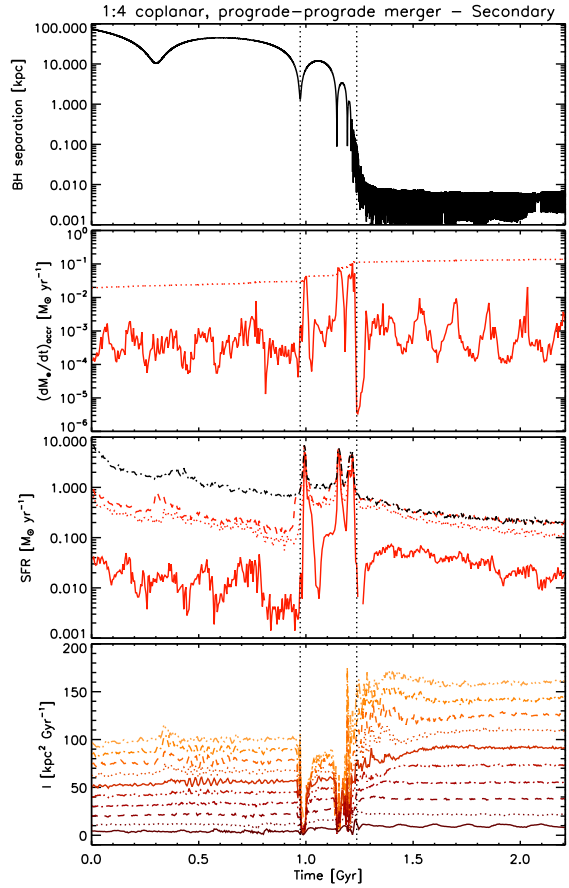}
\vspace{-0.0pt}
\caption[]{Temporal evolution of the main quantities of and around the secondary SMBH of a minor merger from \citet{Capelo_et_al_2015}. In all panels, the two vertical dotted lines show (from left to right) the end of the stochastic stage and the beginning of the remnant stage (see text for more details). First panel: separation between the two SMBHs. Second panel: SMBH accretion rate (solid) and SMBH Eddington accretion rate (dotted). Third panel: global SFR across both galaxies (dash-dotted, black), and SFR in concentric spheres around the SMBH: 0--0.1 (solid, red), 0--1 (dotted, red), and 0--10 (dashed, red) kpc. Fourth panel: gas specific angular momentum magnitude in ten concentric 100-pc shells (0--1~kpc) around the local centre of mass near the SMBH. The colour varies from dark to light orange as the radius of the shell increases. The SMBH accretion rate peaks are correlated with the troughs of gas specific angular momentum. Moreover, nuclear SFR and SMBH accretion rate are proportional to each other \citep[see][]{Volonteri_et_al_2015a,Volonteri_et_al_2015b}. Figure adapted from \citet{Capelo_et_al_2015}, reproduced by permission of Oxford University Press, on behalf of the Royal Astronomical Society.
}
\label{ch1:fig:capelo}
\end{figure*}

Galactic-scale stellar bars \citep[e.g.][]{Sanders_Huntley_1976,Roberts_et_al_1979,Athanassoula_1992} are the most studied example of such mechanisms in the case of isolated systems.\footnote{It is still debated if the origin of bars is internal/secular \citep[e.g.][]{Zana_et_al_2018} or external/tidal \citep[e.g.][]{Romano-Diaz_et_al_2008}. However, the role of bars is more pronounced in isolated galaxies because they are likely the only mechanism that can produce large-scale torques.} However, they can only manage to funnel the gas down to distances of $\sim$0.1--1~kpc, where the bar ceases to be efficient because the gas has reached the inner Lindblad resonance radius \citep[e.g.][]{Binney_Tremaine_2008,Kormendy_2013}. In order to reach the nuclear regions of the galaxy, additional mechanisms are needed, such as systems of nested non-axisymmetric structures \citep[e.g.][]{Hopkins_Quataert_2010}, including nested bars \citep[e.g.][]{Shlosman1989}, or bar-driven nuclear spirals \citep[e.g.][]{Maciejewski_2004,Fanali_et_al_2015}.

Bars are not the only mechanism of note in isolated systems. E.g. dense, massive gas clouds in gas-rich clumpy discs can migrate inwards due to mutual angular momentum exchange with other density perturbations and stochastically collide with the central SMBH \citep[][]{Gabor_Bournaud_2013}, simply travel inwards due to dynamical friction, or move ballistically inwards due to turbulence driven by supernova explosions \citep[][]{Hobbs_et_al_2011}.

Another well-studied mechanism are galaxy mergers, whose dynamics have been investigated since the 1940s \citep[][]{Holmberg_1941}. During the interaction, the tidal field of one galaxy causes the other galaxy to form non-axisymmetric structures which then torque the stars \citep[e.g.][]{Toomre_Toomre_1972} and gas \citep[e.g.][]{Barnes_Hernquist_1996}. Additionally, gas can also lose angular momentum as a consequence of hydrodynamically driven (ram-pressure) shocks, during the closest pericentric passages \citep[][]{Barnes_2002,Capelo_Dotti_2017}: by doing so, gas can effectively become decoupled from the stars and can populate phase-space regions of extremely low angular momentum, possibly unaccessible to the tidal-torque process if, e.g. there exist orbital resonances. The merger-driven gas inflows result in intense bursts of central star formation \citep[e.g.][]{Cox_et_al_2008} and (assuming that the inflow proceeds down to sub-pc scales) SMBH accretion.

The study of SMBH accretion in galaxy mergers is inextricably linked to that of SMBH dynamics (specifically SMBH pairing), since the location of a SMBH can greatly influence the amount of gas available for accretion, especially in the case of the secondary SMBH.

In the past two decades, the dynamics and/or growth of SMBHs have been included in several merger simulations \citep[e.g.][]{DiMatteo_et_al_2005,Hopkins06,Younger_et_al_2008,Johansson_et_al_2009,Callegari09,Callegari_et_al_2011,Hayward_et_al_2014,VanWassenhove_et_al_2014,Capelo_et_al_2015,Gabor_et_al_2016,Pfister_et_al_2017,Khan_et_al_2018}, yielding a scenario exemplified in Figure~\ref{ch1:fig:capelo}, where the evolution of a minor merger from \citet{Capelo_et_al_2015} is shown. The accretion rate on to both the primary and secondary SMBH (capped at a few times the Eddington rate), as well as the SFR, are initially both stochastic and relatively low. When the two galaxies reach the second pericentre (left-hand vertical dotted line), the gas loses much of its angular momentum and both the SMBH accretion rate (especially of the secondary) and SFR reach very high peaks, concurrent to a few pericentric passages. It is during this stage that single and dual \citep[e.g.][]{VanWassenhove_et_al_2012,Blecha_et_al_2013,Capelo_et_al_2017,Blecha_et_al_2017} AGN activity are at their highest. Eventually, a remnant galaxy forms (right-hand vertical dotted line), the gas regains its angular momentum, and SMBH accretion rate and SFR go back to their initial levels.

The SMBHs continue their pairing process in the remnant galaxy, first on galactic scales \citep[e.g.][]{Tamburello_et_al_2017}, then on circumnuclear scales \citep[][]{Dotti_et_al_2007,Fiacconi_et_al_2013,SouzaLima_et_al_2017} and, finally, on sub-pc scales, where they eventually coalesce \citep[e.g.][]{Begelman_et_al_1980}, radiating GWs\footnote{The anisotropic emission of GWs leads to a net emission of linear momentum and to a recoil of the SMBH remnant which, depending on the magnitude and orientation of the spins of the two merging SMBHs, can reach velocities of the order of $10^3$~km~s$^{-1}$ \citep[e.g.][]{Campanelli_et_al_2007}, comparable to the escape velocity of most hosts, thus leading to the possible ejection of the SMBH from its host.} which will be hopefully detected by LISA (e.g. \citealt{eLISA_2013,Klein-goat16,Barack_et_al_2018}; see Chapter~13).

\section{Conclusions}\label{ch1:sec:Conclusions}

A lot has happened since Michell first speculated on the existence of dark stars in 1784. For the past century, we have had a solid theoretical ground (Einstein's GR) to understand BH physics, i.e. the nature and structure of the BHs themselves. We have also had decades of observations across the EM spectrum of sources from all environments and up to $z > 7$, contributing to our understanding of BH astrophysics, i.e. the interaction between these dark objects and the surrounding matter. In 2015, we had the first detection of GWs, letting us listen to the inspiral and coalescence of stellar-mass BHs. Soon, with the current Pulsar Timing Arrays \citep[PTAs;][]{Hobbs_et_al_2010} and planned LISA, we should be able to listen to GW events related to IMBHs and SMBHs.

Despite the enormous advances in BH physics and astrophysics, we still do not know how these objects have formed in the first place. The remainder of this book will be mostly devoted to addressing this fundamental question. We begin laying down the framework in which the first SMBHs formed: how the first dark structures formed (Chapter~2), the chemistry of the primordial gas (Chapter~3), and how the first stars formed (Chapter~4). We then review several modes of SMBH formation --- from the direct collapse of gas (Chapters~5 and 6), stellar clusters (Chapter~7), and supermassive stars (Chapter~8) --- and discuss the statistical predictions from all these methods in Chapter~9. In Chapters 10 and 11, we review both the accretion and feedback of the first SMBHs. Finally, in Chapters~12--14, we describe the current and future observational status, both in the EM and GW domains.\footnote{\tiny Chapter~2: \citet{Chapter2}; Chapter~3: \citet{Chapter3}; Chapter~4: \citet{Chapter4}; Chapter~5: \citet{Chapter5}; Chapter~6: \citet{Chapter6}; Chapter~7: \citet{Chapter7}; Chapter~8: \citet{Chapter8}; Chapter~9: \citet{Chapter9}; Chapter~10: \citet{Chapter10}; Chapter~11: \citet{Chapter11}; Chapter~12: \citet{Chapter12}; Chapter~13: \citet{Chapter13}; Chapter~14: \citet{Chapter14}.\par}

\clearpage

{
\bibliographystyle{ws-rv-har}    
\bibliography{FotFBH_Ch_1}
}


\end{document}